\documentclass[twocolumn]{aastex63}
\usepackage{bm}
\usepackage{blindtext}
\usepackage[utf8]{inputenc}
\usepackage{amsmath}
\newcommand{\bjdtdb}{${\rm {BJD_{TDB}}}$}

\newcommand{\rsun}{${\rm R}_\Sun$}

\newcommand{\mj}{${\,{\rm M}_{\rm J}}$}
\newcommand{\rj}{${\,{\rm R}_{\rm J}}$}

\newcommand{\three}{3.6\,$\mu$m\ }
\newcommand{\four}{4.5\,$\mu$m\ }

\newcommand{\fouralt}{4.5\,$\mu$m}

\begin{document}

\title{The TESS Phase Curve of KELT-1b Suggests a High Dayside Albedo}

\author[0000-0002-9539-4203]{Thomas G. Beatty}
\affiliation{Department of Astronomy and Steward Observatory, University of Arizona, Tucson, AZ 85721; tgbeatty@arizona.edu}

\author[0000-0001-9665-8429]{Ian Wong}
\altaffiliation{51 Pegasi b Fellow}
\affiliation{Department of Earth, Atmospheric, and Planetary Sciences, Massachusetts Institute of Technology, Cambridge, MA 02139}

\author[0000-0002-3551-279X]{Tara Fetherolf}
\affiliation{Department of Physics and Astronomy, University of California, Riverside, CA 92521}

\author[0000-0002-2338-476X]{Michael R. Line}
\affiliation{School of Earth \& Space Exploration, Arizona State University, Tempe, AZ 85287}

\author[0000-0002-1836-3120]{Avi Shporer}
\affiliation{Department of Physics and Kavli Institute for Astrophysics and Space Research, Massachusetts Institute of Technology, Cambridge, MA 02139}

\author[0000-0002-3481-9052]{Keivan G. Stassun}
\affiliation{Vanderbilt University, Department of Physics \& Astronomy, 6301 Stevenson Center Lane, Nashville, TN 37235}
\affiliation{Fisk University, Department of Physics, 1000 17th Avenue N., Nashville, TN 37208}

\author[0000-0003-2058-6662]{George R. Ricker}
\affiliation{Department of Physics and Kavli Institute for Astrophysics and Space Research, Massachusetts Institute of Technology, Cambridge, MA 02139, USA}

\author[0000-0002-6892-6948]{Sara Seager}
\affiliation{Department of Physics and Kavli Institute for Astrophysics and Space Research, Massachusetts Institute of Technology, Cambridge, MA 02139, USA}
\affiliation{Department of Earth, Atmospheric and Planetary Sciences, Massachusetts Institute of Technology, Cambridge, MA 02139, USA}
\affiliation{Department of Aeronautics and Astronautics, MIT, 77 Massachusetts Avenue, Cambridge, MA 02139, USA}

\author[0000-0002-4265-047X]{Joshua N. Winn}
\affiliation{Department of Astrophysical Sciences, Princeton University, Princeton, NJ 08544, USA}

\author[0000-0002-4715-9460]{Jon M. Jenkins}
\affiliation{NASA Ames Research Center, Moffett Field, CA 94035, USA}

\author[0000-0002-2457-272X]{Dana R. Louie}
\affiliation{Department of Astronomy, University of Maryland, College Park, MD 20742, USA}

\author[0000-0001-5347-7062]{Joshua E. Schlieder}
\affiliation{NASA Goddard Space Flight Center, Greenbelt, MD 20771, USA}	

\author[0000-0001-5401-8079]{Lizhou Sha}
\affiliation{Department of Physics and Kavli Institute for Astrophysics and Space Research, Massachusetts Institute of Technology, Cambridge, MA 02139, USA}

\author[0000-0002-1949-4720]{Peter Tenenbaum}
\affiliation{SETI Institute / NASA Ames Research Center, Moffett Field, CA 94035, USA}

\author[0000-0003-4755-584X]{Daniel A. Yahalomi}
\affiliation{Center for Astrophysics ${\rm \mid}$ Harvard {\rm \&} Smithsonian, 60 Garden Street, Cambridge, MA 02138, USA}


\begin{abstract}
We measured the optical phase curve of the transiting brown dwarf KELT-1b \citep[TOI 1476,][]{siverd2012} using data from the TESS spacecraft. We found that KELT-1b shows significant phase variation in the TESS bandpass, with a relatively large phase amplitude of $234^{+43}_{-44}$\,ppm and a secondary eclipse depth of $371^{+47}_{-49}$\,ppm. We also measured a marginal eastward offset in the dayside hotspot of $18.3^\circ\pm7.4^\circ$ relative to the substellar point. We detected a strong phase curve signal attributed to ellipsoidal distortion of the host star, with an amplitude of $399\pm19$\,ppm. Our results are roughly consistent with the Spitzer phase curves of KELT-1b \citep{beatty2019}, but the TESS eclipse depth is deeper than expected. Our cloud-free 1D models of KELT-1b's dayside emission are unable to fit the full combined eclipse spectrum. Instead, the large TESS eclipse depth suggests that KELT-1b may have a significant dayside geometric albedo of $\mathrm{A}_\mathrm{g}\sim0.5$ in the TESS bandpass, which would agree with the tentative trend between equilibrium temperature and geometric albedo recently suggested by \cite{wong2020}. We posit that if KELT-1b has a high dayside albedo, it is likely due to silicate clouds \citep{gao2020} that form on KELT-1b's nightside \citep{beatty2019,keating2019} and are subsequently transported onto the western side of KELT-1b's dayside hemisphere before breaking up.
\end{abstract}

\section{Introduction}

Optical phase curve observations of hot Jupiters have been relatively rare. Though many of these planets have had their phase curves observed in the near infrared (NIR), until recently there were only 15 hot Jupiters around bright stars with optical phase curve data -- all from Kepler \citep{angerhausen2015,esteves2015}. The TESS mission has dramatically changed this picture, and there are now 13 additional systems with published results \citep{shporer2019,bourrier2020,wong2020,wong2020kelt9,wong2020wasp19,vonessen2020} -- with many more on the way as TESS completes its two-year Primary Mission.

The increased number of hot Jupiters with precise optical phase curve data allows us to study in more detail the emission properties and the geometric albedos of hot giant planet atmospheres at these wavelengths \citep{mayorga2019}. Albedo measurements are particularly sensitive to the presence of clouds in the atmosphere, giving us information about their composition \citep[e.g.,][]{oreshenko2016,parmentier2016}. When combined with a good understanding of the thermal properties of the atmosphere, optical phase curve measurements also provide a rough picture of the distribution of clouds across the planet's surface \citep{demory2013, shporer2015}.

Optical eclipse observations of hot and ultra-hot Jupiters have shown that the integrated daysides of these planets typically do not have high geometric albedos, and hence not much reflective cloud cover. Ensemble analyses of Kepler observations indicate that the range of geometric albedos for hot Jupiters is $A_g\lesssim0.2$ \citep[][]{angerhausen2015, esteves2015}, though these results rely on assumptions about the exact amount of contaminating thermal emission in the broad Kepler bandpass. Albedo measurements solely in the blue-optical are one method to avoid this thermal contamination, and HST/STIS eclipse observations of WASP-12b from 290\,nm to 570\,nm by \cite{bell2017} showed no detectable signal. This implies an upper limit of $A_g<0.064$ for the planet's geometric albedo in this wavelength range and raises the possibility that unmodeled thermal emission is artificially raising the measured albedos in the Kepler studies. However, HD 189733b does show a clear STIS eclipse signal from 290\,nm to 450\,nm \citep[$A_g=0.40\pm0.12$,][]{evans2013}. Complicating the picture, HD 189733b's 450\,nm to 580\,nm eclipse is consistent with zero. HD 189733b has an equilibrium temperature approximately 1500\,K colder than WASP-12b and other ultra-hot Jupiters, and therefore has dayside temperatures much more amenable to cloud formation.

Indeed, 3D global circulation models (GCMs) of exoplanet atmospheres typically show that the daysides of ultra-hot Jupiters are too hot for clouds to form, except in a narrow region near the western terminator \citep[e.g.,][]{parmentier2016,powell2018,helling2019}. However, these studies do not include the effects of horizontal transport on their cloud distributions, nor the possible effects of cloud feedback and self-shielding. Properly incorporating these effects -- particularly horizontal transport -- into 3D cloud models may significantly change the expected range of dayside clouds on these planets \cite[e.g.,][]{helling2019}. 

Recently, \cite{wong2020} conducted an ensemble analysis of TESS and Spitzer secondary eclipse depths to better constrain the geometric albedos of hot Jupiters. Using the long wavelength Spitzer results, \cite{wong2020} were able to model each planet's thermal emission, rather than being forced to assume it as in \cite{angerhausen2015} and \cite{esteves2015}, which allowed them to more reliably measure each planet's albedo in the TESS bandpass. They found average geometric albedos in the TESS bandpass of $A_g\approx0.2$ for their sample. \cite{wong2020} also found a tentative  correlation between increasing temperature and increasing geometric albedo for planets with dayside temperatures between 1500\,K and 3000\,K. This more detailed analysis corroborates the albedo trends reported in \cite{angerhausen2015} and \cite{esteves2015} and suggests that the hottest planets may in fact have systematically enhanced apparent albedos in the red-optical (i.e., about 600\,nm to 1,000\,nm). Beyond 3000\,K, the supermassive ultra-hot Jupiter WASP-18b has a geometric albedo consistent with zero, marking an apparent break in the trend. More albedo measurements for planets in this temperature range are needed to ascertain whether this transition is related to a temperature-dependent process in the atmospheric chemistry or a consequence of the significantly higher surface gravity of WASP-18b relative to lower-mass hot Jupiters.

\begin{figure*}[!ht]
    \centering
    \includegraphics[width=\linewidth]{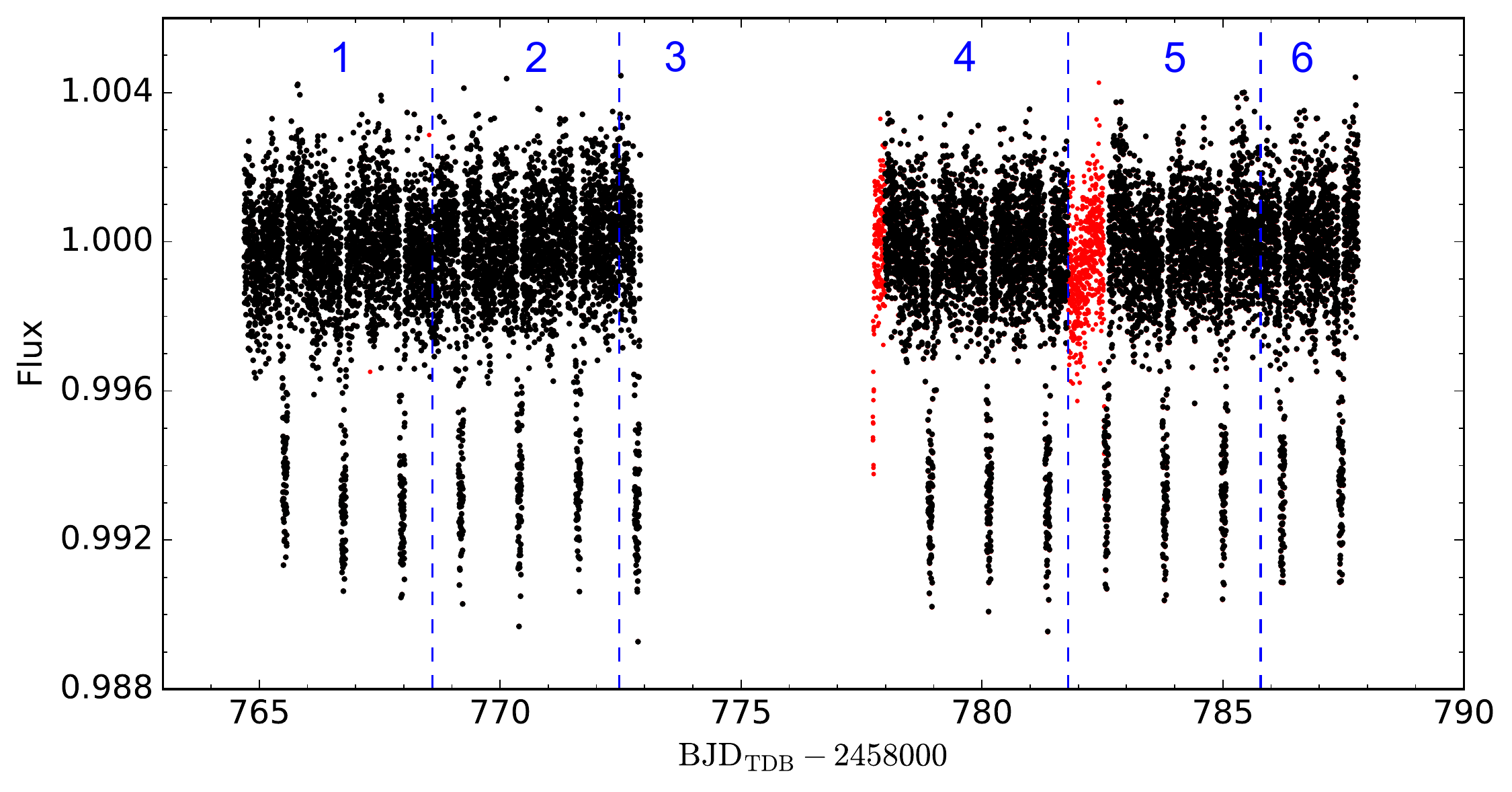}
    \caption{Normalized outlier-removed TESS PDC light curve of KELT-1. The momentum dumps are indicated by the vertical blue lines. The data segments are labeled 1--6. The red points denote the trimmed flux ramps at the beginning of segments 4 and 5.}
    \label{fig:lc}
\end{figure*}

Observations of the transiting brown dwarf KELT-1b give us another object with which to test this emerging trend. KELT-1b is a 27.23\mj\ brown dwarf with a radius of 1.1\rj\ that is on a 1.27 day orbit around its parent star \citep{siverd2012}. The dayside temperature measured from Spitzer observations of KELT-1b is 2950\,K \citep{beatty2019}, which is similar to ultra-hot Jupiters. Unlike ultra-hot Jupiters, however, KELT-1b has a high surface gravity of $\log\,g=4.75$, nearly 30 times higher than the typical ultra-hot Jupiter.

Secondary eclipses of KELT-1b have been previously measured from the ground at $z^\prime$ \citep{beatty2014}, $H$ \citep{beatty2017}, $K_s$ \citep{croll2015}, and there are existing phase curve observations from Spitzer at \three and \fouralt\ \citep{beatty2019}. This gives us several measurements of the thermal emission from KELT-1b's atmosphere and should in principle allow for the contribution of thermal emission to be subtracted from the object's brightness in the TESS bandpass and provide constraints on the dayside geometric albedo.

\section{Data Analysis}\label{section:data}

TESS observed the KELT-1 system ($V$ = 10.63 mag, Tmag = 10.22) from 2019 Oct 7 to 2019 Nov 2 during its Sector 17 observations. KELT-1 was one of the targets pre-selected for 2 minute cadence observations by TESS as TOI 1476.01 and is listed in the TESS Input Catalog (TIC) as TIC 432549364. TESS collected a total of 18,012 flux measurements of KELT-1.

We used the Pre-search Data Conditioning \citep[PDC,][]{pdc1, pdc2} light curve as the basis for our data analysis. These data were reduced using the Science Processing Operations Center (SPOC) pipeline \citep{spoc}, which determined the optimal extraction aperture within the $11\times11$ pixel ``postage stamp'' centered on KELT-1 and applied corrections for instrumental systematics to the extracted light curve.

During the observations of each sector, the TESS spacecraft completes two 13.7-day elliptic orbits around the Earth, with interruptions in data collection occurring during perigee for data downlink. TESS Sector 17 observations were affected by several episodes of Earth scattered light on the CCD. These occurred during the last 3.5 and 2.5 days of the first and second orbits of Sector 17, respectively. The SPOC pipeline marked these exposures with quality flags, indicating bad quality data, along with all other data points that may have been affected by cosmic rays or other non-nominal spacecraft operation. We removed all 4,872 flagged data points from the time series.

Momentum dumps are scheduled throughout each spacecraft orbit to reset the reaction wheels. During Sector 17 observations, these occurred twice per orbit. Often, there are discontinuities in the light curve across these events, as well as additional flux ramps immediately preceding or following the momentum dump. Because such short-timescale photometric variations are difficult to correct for without fitting away part of the astrophysical phase curve variation, we trimmed them from the light curve prior to analysis. In the KELT-1 dataset (Figure \ref{fig:lc}), significant ramps occurred at the beginning of the second spacecraft orbit and following the subsequent momentum dump, and we trimmed 0.25 and 0.75 days worth of data during those instances, respectively.

After removing the data points near the momentum dumps, we applied a 16 point wide $3\sigma$ moving median filter to the transit- and eclipse-masked light curve. The final trimmed and outlier-removed KELT-1 light curve for our analysis contains 12,287 data points. The median fractional uncertainty in each measurement was 1247 ppm. The light curve is plotted in Figure \ref{fig:lc}, with the times of momentum dumps indicated by vertical lines and trimmed ramps denoted by the red points.

\subsection{Phase Curve Model}\label{section:fitmodel}

The phase curve model that we used to fit the TESS lightcurve of KELT-1b included models for the transits, eclipses, and phase variation, as well as signals from the ellipsoidal deformation and Doppler boosting of the host star (for a review see \citealt{shporer2017}). Due to the relatively high mass of KELT-1b (27.2\mj), the amplitude of the latter two terms are significantly larger than in the case of a typical hot Jupiter.

Both our transit and eclipse models used the \textsc{batman} package \citep{batman}, which is a Python implementation of the \cite{mandel2002} lightcurve model. The free parameters in both models were the transit center time ($T_C$), the orbital period (as $\log P$), the orbital inclination (as $\cos i$), the star-companion radius ratio ($R_{BD}/R_*$), and the scaled semi-major axis of the orbit (as $\log a/R_*$). We set the secondary eclipse depth based on the phase curve parameters described below. We calculated the secondary eclipse time using the transit center time and the orbital period assuming a circular orbit. We included a delay in the eclipse time to account for light travel time across KELT-1b's orbit, for a given value of $a/R_*$ and assuming $R_*=1.482$\,\rsun\ (see Section 2.3).

For limb darkening, we used the standard quadratic limb darkening law and fixed the coefficients to the tabulated values in \citet{claret2018} for the nearest combination of stellar parameters: $u_1=0.3287$, $u_2=0.2160$.

All seven of the free parameters have been measured in previous observations, and we used these independent measurements and their associated uncertainties as Gaussian priors in our fitting process. Specifically, we used the results from the Spitzer phase curve observations of KELT-1b \citep{beatty2019}, which we list for reference in Table \ref{tab:priors}.

Our model for KELT-1b's atmospheric brightness variation is a single sinusoid with a variable amplitude $C_1$, phase offset $C_2$, and zero-point $F_{0}$,
\begin{equation}\label{eq:3210}
F_{atm}(\phi) = F_0 - C_1\,\cos\left(\frac{2\pi(t-T_C)}{P}+C_2\right).
\end{equation}
We imposed no priors on any of the phase curve parameters. 

The mutual gravitational interaction between the host star and the brown dwarf imparts time-varying signals to the star's flux \citep[e.g.,][]{shporer2017}. The ellipsoidal distortion modulation stems from the tidal bulge raised on the host star by the orbiting companion, and the leading term in this photometric signal takes the form of a cosine at the first harmonic of the orbital phase $\phi$ \citep[e.g.,][]{beatty2019,koi964}:
\begin{equation}
    F_{ellip} = -A_{ellip}\cos\left(4\pi\phi \right).\label{ellip}
\end{equation}
Here, the average stellar brightness is normalized to unity, and the amplitude $A_{ellip}$ depends on the star-companion mass ratio and other fundamental orbital parameters \citep[e.g.,][]{mazeh2010,koi964}:
\begin{equation}
    A_{ellip}= \beta \frac{M_{BD}}{M_{*}}\left( \frac{R_{*}}{a} \right)^{3} \sin^{2} i,\label{ellipamp}
\end{equation}
where the prefactor $\beta$ depends on the linear limb-darkening and gravity-darkening coefficients of the star. 

The second of the gravitationally induced signals is Doppler boosting, which arises from the relativistic beaming and Doppler shifting of the star's emission due to the radial velocity induced by the orbiting companion. The amplitude of this signal is given by \citep[e.g.,][]{loeb2003}
\begin{equation}
    F_{Dopp} =  A_{Dopp}\sin\left(2\pi\phi \right),\label{doppler}
\end{equation}
where the amplitude $A_{Dopp}$ is related to the orbital radial velocity semi-amplitude $K_{RV}$ via
\begin{equation}
    A_{Dopp} = \alpha\left( \frac{K_{RV}}{c} \right).\label{doppleramp}
\end{equation}
The prefactor $\alpha$ depends on the shape of the stellar spectrum and the observed wavelength. 

Following the methods detailed in \cite{koi964} and the expressions described above, we can calculate the predicted ellipsoidal distortion and Doppler boosting amplitudes. To compute the prefactor values, we fit 3D polynomials to determine the limb-darkening, gravity-darkening, and $\alpha$ values of KELT-1 as a function of stellar effective temperature, metallicity, and surface gravity. For both the limb-darkening and gravity-darkening coefficients, we linearly interpolated between the tabulated values from \cite{claret2017}, and for $\alpha$, we directly interpolated the value from a grid of \textsc{phoenix} stellar models \citep{husser2013}. 

\begin{deluxetable}{lcl}[t]
\tablecaption{Prior Values for KELT-1b's Properties\\ From \cite{beatty2019}}
\tablehead{\colhead{Parameter} & \colhead{Units} & \colhead{Value}}
\startdata
$T_C$\dotfill &Transit time (\bjdtdb)\dotfill & $2457306.97602\pm0.0003$\\
$P$\dotfill &Orbital period (days)\dotfill & $1.2174928\pm6\times10^{-7}$\\
$\sqrt{e}\cos{\omega}$\dotfill & \dotfill & $\equiv 0$\tablenotemark{a}\\
$\sqrt{e}\sin{\omega}$\dotfill & \dotfill & $\equiv 0$\tablenotemark{a}\\
$\cos{i}$\dotfill &Cosine of inclination\dotfill & $0.054\pm0.015$\\
$R_{BD}/R_{*}$\dotfill &Radius ratio\dotfill & $0.0771\pm0.0003$\\
$a/R_{*}$\dotfill &Scaled semimajor axis\dotfill & $3.693\pm0.038$\\
$M_{BD}/M_{*}$\dotfill & Mass ratio\dotfill & $0.01958\pm0.0004$\\
\enddata
\tablenotetext{a}{\cite{beatty2019} found both these consistent with zero at $<1\,\sigma$. For simplicity we therefore assumed a circular orbit.}
\label{tab:priors}
\vskip -0.2in
\end{deluxetable}

Propagating the uncertainties in the stellar parameters forward via Monte Carlo sampling, we arrived at a predicted Doppler boosting amplitude of $A_{Dopp} = 41.1\pm1.0$~ppm. For ellipsoidal distortion, we utilized literature values for the mass ratio and other system properties (see Table~\ref{tab:priors}) to arrive at a predicted amplitude of $A_{ellip}=417\pm26$~ppm. We note that the ellipsoidal distortion signal is formally an expansion of cosine terms, with the second-order term situated at the second harmonic of the orbital period --- in other words, $\cos(6\pi\phi)$. Using the same estimation method and the formalism described in \citet{koi964}, we found that the second-order amplitude is $32\pm5$~ppm.

Given the precise prior information we have about KELT-1's physical properties and the radial velocity orbit of the system, we placed a Gaussian prior on the Doppler boosting amplitude based on our aforementioned estimate when fitting the TESS light curve. Meanwhile, we wanted to empirically test whether the measured ellipsoidal distortion amplitude is consistent with the predictions of theory (see \citealt{wong2020kelt9} and \citealt{koi964} for an in-depth discussion of various caveats and assumptions inherent within the classical theory of stellar tidal response). Therefore, in our analysis, we fit for the ellipsoidal distortion signal by allowing either the amplitude $A_{ellip}$ (in the case of polynomial detrending; Section~\ref{section:polyfit}) or the mass ratio $M_{BD}/M_{*}$ (in the case of Gaussian-process regression; Section~\ref{section:gp}) to vary freely. We experimented with measuring the second-order ellipsoidal distortion signal at the second harmonic, but did not detect any significant amplitude; in the fits presented in this paper, we did not include this higher-order modulation in the phase curve modeling.

The combined out-of-eclipse phase curve model, normalized such that the average combined brightness of the star and brown dwarf is unity, is
\begin{equation}
    F(\phi) = \frac{1+F_{atm}+F_{ellip}+F_{Dopp}}{1+F_{0}}.\label{full}
\end{equation}

\subsection{Systematics Detrending}\label{section:systematics}

Some residual systematic trends not removed by the SPOC pipeline are discernible in the KELT-1 light curve (Figure~\ref{fig:lc}). To fit the phase curve signals while accounting for these background trends, we used three different detrending methods: polynomial detrending, Gaussian-Process regression, and Fourier decomposition of the transit- and eclipse-removed light curve. All three methods produced a consistent set of astrophysical parameters, and we chose the fitted parameters from the polynomial detrending method as the primary results of this work.

\subsubsection{Polynomial Detrending}\label{section:polyfit}

This systematics detrending method has been utilized in several previously published TESS phase curve analyses \citep{shporer2019,wong2020kelt9,wong2020wasp19}. Using the ExoTEP pipeline (see, for example, \citealt{benneke2019}), we divided the KELT-1 light curve into six segments separated by the momentum dumps. For each segment $i$, we multiplied the astrophysical phase curve model by a generalized polynomial in time of order $N$:
\begin{equation}\label{systematics}
    S_N^{\lbrace i\rbrace}(t) = \sum\limits_{j=0}^{N}c_j^{\lbrace i\rbrace}(t-t_0)^j.
\end{equation}
Here, $t_0$ is the time of the first data point in the segment. 

To determine the optimal polynomial orders, we first carried out individual fits of each segment using polynomial detrending models of different orders. We then considered both the Akaike Information Criterion ($\mathrm{AIC}\equiv 2\gamma -2 \log L$) and the Bayesian Information Criterion ($\mathrm{BIC}\equiv \gamma\log m -2 \log L$); $\gamma$ is the number of free parameters in the fit, $m$ is the number of points in the fitted dataset, and $L$ is the maximum log-likelihood. Both of these statistical metrics balance improvements to the model fitting from increasing the order of the detrending polynomial(s) with penalties for the inclusion of additional detrending parameters; the BIC penalizes extra parameters more strongly than the AIC. By minimizing the AIC, we found that the optimal polynomial orders for the six segments are 6, 5, 0, 7, 6, and 3; doing the same with the BIC instead yielded 6, 1, 0, 3, 4, and 1.

When using polynomials to correct for systematics in a time series, biases in the fitted parameters can arise when the number of inflection points in the detrending model is near to or exceeds the number of inflection points in the underlying astrophysical signal. This is evidenced by the emergence of covariances between the polynomial coefficients and the phase curve amplitudes. Such covariances occurred in several of the data segments for which the AIC prefers a high-order polynomial. However, when comparing the results for the cases in which the AIC and BIC prefer different polynomial orders, we found that the phase curve amplitudes always agree to within $2\sigma$. Moreover, when combining the segments together into a joint fit, the covariances between the individual systematics coefficients and the phase curve amplitudes are greatly reduced, since the astrophysical signal is shared across all segments.

To determine the final set of optimal polynomial orders, we carried out joint fits of all six segments, using all possible combinations of polynomial orders preferred by the AIC or BIC from the individual segment fits. We then chose the combination of orders that minimized the overall BIC from the joint fit to use in generating the final results: 6, 1, 0, 7, 6, and 1. Across the various joint fits with different combinations of polynomial orders, the fitted phase curve amplitudes were self-consistent to well within $1.5\sigma$, indicating that the specific choice of polynomial orders does not have any significant effect on the results. In addition, for the joint fit using our final choice of polynomial orders, there are no significant covariances between polynomial coefficients and phase curve amplitudes.

\subsubsection{Gaussian-Process Regression}\label{section:gp}

We also fit KELT-1b's phase curve using a Gaussian-Process (GP) regression to fit for the remaining systematics trends in the data. GP regressions have been used before to fit systematics in TESS phase curves \citep{daylan2019}, and their general use for detrending photometric observations is more fully described in \cite{gibson2012}. We used the same astrophysical model as in Section 2.2.1, with the same priors on the KELT-1 system parameters.

Due to the relatively large size of the TESS light curve, it was numerically impossible to perform the necessary matrix inversions required to compute a GP model using a typical covariance kernel (e.g., a squared exponential). Instead, we used the \textsc{celerite} Python package \citep{celerite} to conduct our GP regression. \textsc{celerite} restricts the choice of possible covariance kernels to those that yield invertable covariance matrices in $\mathcal{O}(n)$ time. We chose to use an exponential kernel for our GP regression.

One consideration in using GP regression to fit the light curves of transiting exoplanets is ensuring that the length scales of the GP covariances are not allowed to run shorter than the scale of the astrophysical signal \citep{beatty2017}. A GP regression with such a short length scale will begin to fit the astrophysical signal itself, and resulting parameter uncertainties will be large. \textsc{celerite}'s exponential kernel parameterizes this using the logarithm of the inverse of the covariance length, $\log(c)$, where $c$ is the covariance length inverse. In the case of our fit, this translated to a requirement that $\log(c)<-12$. This forced the GP regression to longer length scales.

We fit the light curve using the GP regression by performing an initial Nelder-Mead likelihood maximization, followed by an MCMC fit about that maximum to determine parameter uncertainties and the true global likelihood maximum. As mentioned in Section 2.1, we used Gaussian priors on the system parameters listed in Table~\ref{tab:priors}. We used the \textsc{emcee} Python package \citep{emcee} to conduct the MCMC fit. We ran the MCMC fit for an initial burn-in of 24,000 steps, followed by a production run of 48,000 steps. We checked the convergence of the MCMC chains by verifying that the Gelman-Rubin statistics of all the chains were below 1.1.

The results from our GP regression fit to KELT-1b's phase curve are described in Section 3 and listed in the fourth column of Table \ref{tab:results}. All of the parameters are consistent with the results from fit using the polynomial detrending method, though the GP phase curve parameters have larger uncertainties. This is an expected property of systematics detrending using a GP regression \citep[e.g.][]{beatty2017}.

\subsubsection{Out-of-Transit Fit}

As a quick sanity check, we also measured KELT-1b's phase curve using a simplified model that was fit to only the out-of-eclipse portions of the light curve. We assumed a circular orbit ($e=0$) and used the orbital parameters (transit time $T_C$, orbital period $P$, and transit duration $T_{14}$) from \citet{beatty2017} and \citet{beatty2019} to remove the expected transits and occultations of KELT-1b. For these fits, we used a least-squares minimization technique to derive the best-fit coefficients. 

In addition to the data processing described in Section~\ref{section:data}, for this analysis we split the light curve into two halves (corresponding to each TESS orbit) and detrended each half by the best-fit linear trend. Since KELT-1 is a fairly quiet star, there were negligible changes in the results when detrending by higher order polynomials ($N\leq$10). We then iteratively fit the phase curve in order to remove all outliers that were greater than 4.5$\sigma$. 

Using a basic Fourier decomposition routine, we measured the best-fit amplitudes of the $\cos(2\pi\phi)$, $\sin(2\pi\phi)$, and $\cos(4\pi\phi)$ harmonic terms: $186$\,ppm, $100$\,ppm, and $415$\,ppm, respectively. The third amplitude corresponds to the ellipsoidal distortion of the host star. In order to disentangle the shift in the atmospheric brightness modulation of KELT-1b, we used the predicted Doppler boosting amplitude of $41.1$\,ppm and removed the expected Doppler modulation from the combined measured phase curve signal at the fundamental of the orbital period ($\sin(2\pi\phi)$ and $\cos(2\pi\phi)$).

Through this process, we inferred that the shifted atmospheric phase curve signal has a semi-amplitude of 195\,ppm and a dayside brightness maximum roughly $18^{\circ}$ east of the substellar point. These results are broadly consistent with the results of the GP regression analysis presented in Section~\ref{section:gp} and the adopted polynomial detrending fit described in Section~\ref{section:polyfit}. We note that amplitude of the atmospheric brightness modulation is most susceptible to biases upon the removal of the transits and secondary eclipses, since the trimming takes away points near both the maximum and minimum of the characteristic signal from the light curve.

\subsection{Broadband Spectral Energy Distribution}\label{section:sedfit}

\begin{figure}[t]
\centering
\includegraphics[angle=180,width=\linewidth,trim=70 50 140 70,clip]{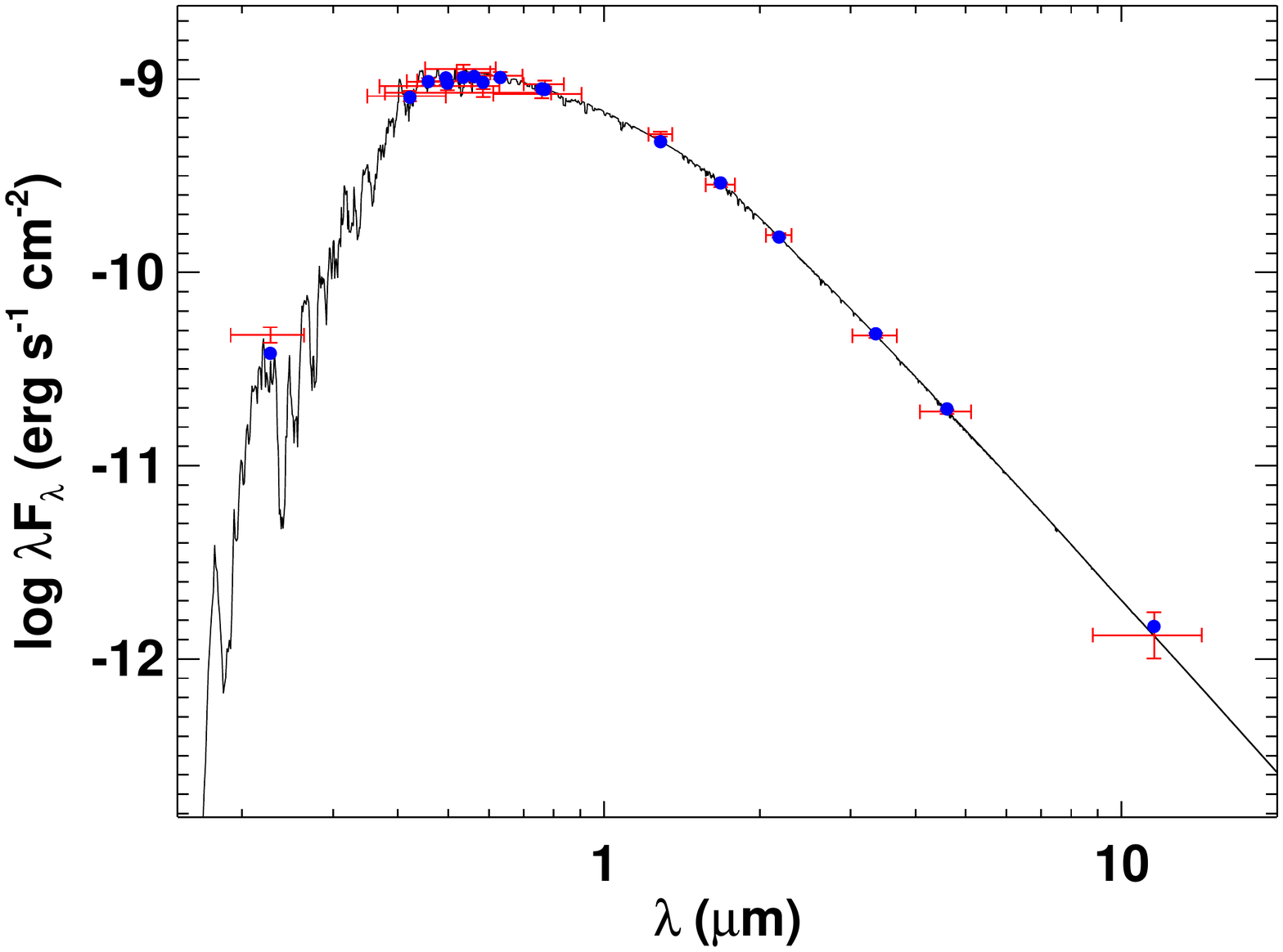}
\caption{Spectral energy distribution (SED). Red symbols represent the observed photometric measurements, where the horizontal bars represent the effective width of the passband. Blue symbols are the model fluxes from the best-fit Kurucz atmosphere model (black).}
\label{fig:sed}
\end{figure}

With the availability of {\it Gaia\/} DR2 parallax and photometry, which were not available at the time of KELT-1b's discovery publication, we performed an updated spectral energy distribution (SED) analysis as an independent check on the derived stellar parameters. Here, we used the SED together with the {\it Gaia\/} DR2 parallax in order to estimate the effective temperature ($T_{\rm eff}$) and determine an empirical measurement of the stellar radius following the procedures described in \citet{StassunTorres2016,Stassun2017,Stassun2018}. We pulled the $B_T V_T$ magnitudes from Tycho-2, the $BVgri$ magnitudes from APASS, the $JHK_S$ magnitudes from {\it 2MASS}, the W1--W3 magnitudes from {\it WISE}, the $G G_{\rm BP} G_{\rm RP}$ magnitudes from {\it Gaia}, and the NUV magnitude from {\it GALEX}. Together, the available photometry spans the stellar SED over the wavelength range 0.2--10~$\mu$m (see Figure~\ref{fig:sed}). 

We performed a fit using \textsc{phoenix} stellar atmosphere models \citep{husser2013}, with the principal free parameters being $T_{\rm eff}$ and the extinction ($A_V$), which we restricted to the maximum line-of-sight value from the dust maps of \citet{Schlegel1998}. The SED does not strongly constrain surface gravity ($\log g$) or metallicity ([Fe/H]), and hence we adopted the values from the discovery paper. The resulting fit is excellent (Figure~\ref{fig:sed}) with a reduced $\chi^2$ of 1.6. The best-fit temperature is $T_{\rm eff} = 6500\pm 150$~K, and the best-fit extinction is $A_V = 0.25\pm 0.04$. Integrating the (unreddened) model SED gives the bolometric flux at Earth of $F_{\rm bol} = 1.621 \pm 0.057 \times 10^{-9}$ erg~s$^{-1}$~cm$^{-2}$. Taking the $F_{\rm bol}$ and $T_{\rm eff}$ together with the {\it Gaia\/} DR2 parallax, adjusted by $+0.08$~mas to account for the systematic offset reported by \citet{StassunTorres2018}, we computed an updated stellar radius of $R = 1.482 \pm 0.070$~R$_\odot$. 

\section{Results}

\begin{figure}[t]
\centering
\includegraphics[width=\linewidth]{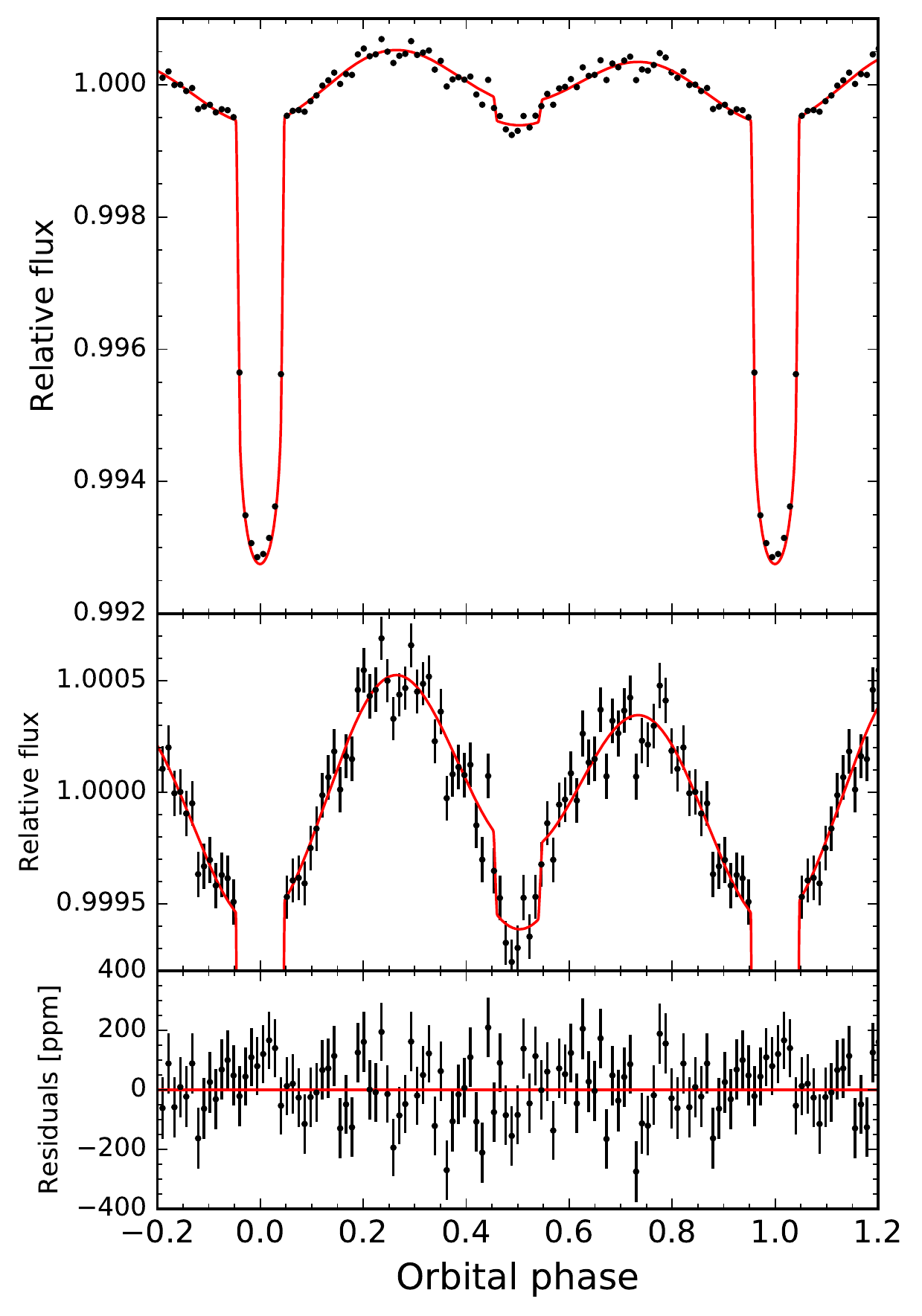}
\caption{Top panel: systematics-removed, phase-folded light curve of KELT-1, binned in 20-minute intervals (black points), alongside the best-fit full phase curve model (red curve). Middle panel: zoomed in view of the phase curve variations and secondary eclipse. Bottom panel: corresponding residuals from the best-fit model.}
\label{fig:fit}
\end{figure}

\begin{figure}[t]
\centering
\includegraphics[width=\linewidth]{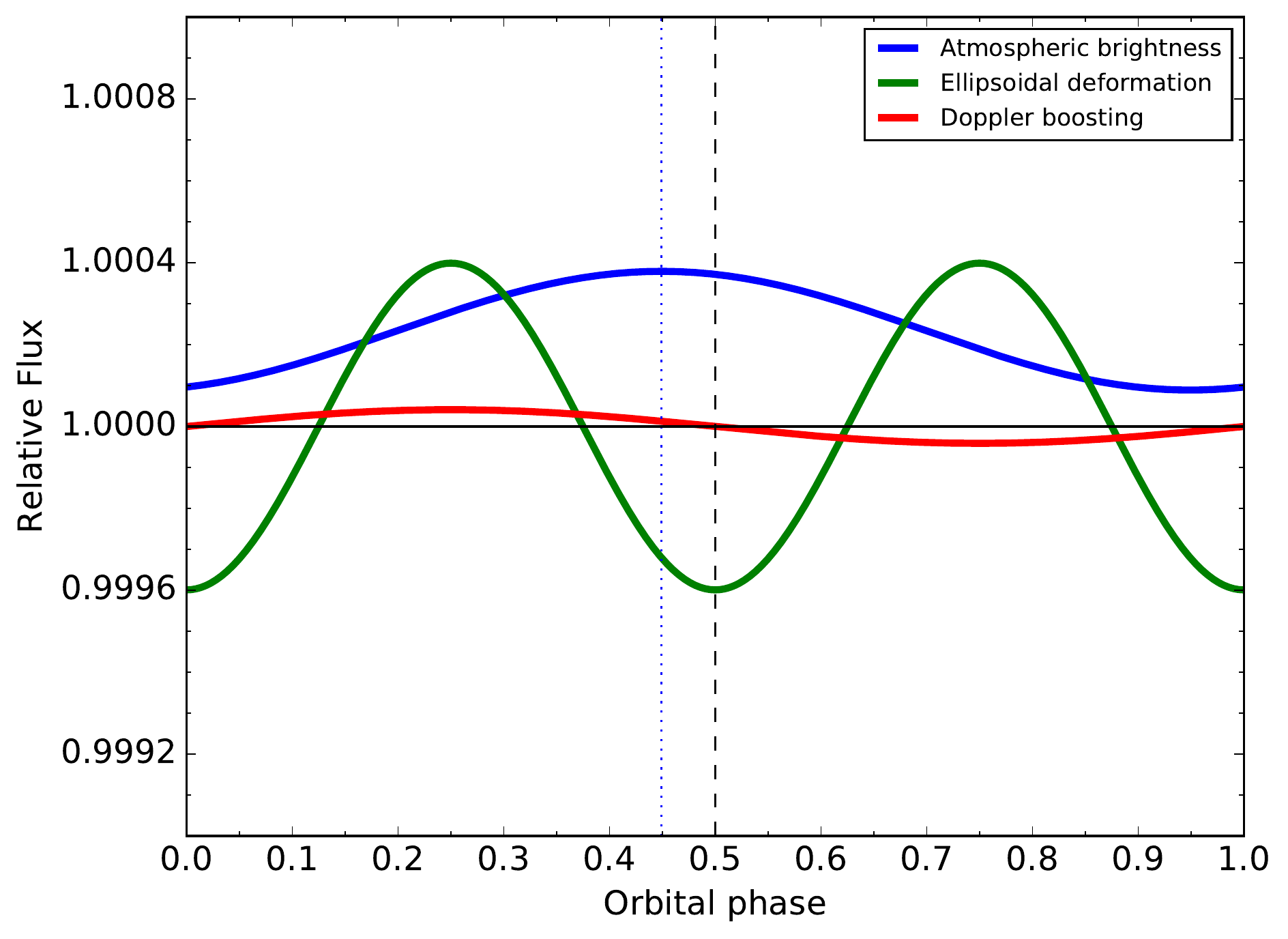}
\caption{Diagram showing the fitted three phase curve components in the KELT-1 light curve. The flux is normalized such that the average brightness of the host star is unity. The atmospheric brightness modulation of the brown dwarf is plotted with the blue curve; the vertical blue dashed line indicates the phase of maximum brown dwarf brightness, which is shifted relative to mid-orbit (vertical black line). The green and red curves show the ellipsoidal deformation and Doppler boosting components of the host star's photometric modulation.}
\label{fig:components}
\end{figure}

We adopted the results from polynomial detrending as the best fit values for KELT-1b's phase curve, though all three analyses provided consistent results (Table 2). We find that KELT-1b shows a relatively large atmospheric phase curve amplitude of $234^{+43}_{-44}$\,ppm and a mildly significant phase offset of $18.3^\circ\pm7.4^\circ$. KELT-1b's secondary eclipse depth is also relatively large, at $371^{+47}_{-49}$\,ppm, while the nightside flux level is not significantly above zero. The systematics-removed, phase-folded, and binned light curve is shown in Figure~\ref{fig:fit}, alongside the best-fit phase curve model. The individual components of the phase curve signal are plotted separately in Figure~\ref{fig:components}.

The Doppler boosting amplitude was constrained by Gaussian priors, and we obtained $41.1\pm1.0$\,ppm. Meanwhile, the ellipsoidal distortion signal was unconstrained in our fit, and we measured an amplitude of $399\pm19$\,ppm, consistent with the theoretical prediction of $417\pm26$~ppm to well within the $1\sigma$ level. This indicates that the photometric signal stemming from the tidal distortion of the host star is well-described by the physical formalism described in Equations~\eqref{ellip}--\eqref{ellipamp}.

To calculate the blackbody brightness temperatures in the TESS bandpass, we used a \textsc{phoenix} model spectrum \cite{husser2013} for the host star derived using the stellar parameters from Section 2.3 and estimated the uncertainties using MCMC, following \citep{beatty2019}. The dayside of KELT-1b has a blackbody brightness temperature of $3340\pm110$\,K, while the nightside has a poorly constrained brightness temperature of $1820^{+640}_{-1150}$\,K.

\section{Discussion}

The overall system and phase curve parameters from the TESS data are generally consistent with results from the previous Spitzer phase curve observations of KELT-1b \citep{beatty2019}. In particular, the phase curve offset in the TESS band ($18.3^\circ\pm7.4^\circ$) is similar to the measured phase offsets at \three ($28.4^\circ\pm3.5^\circ$) and \four ($18.6^\circ\pm5.2^\circ$). This makes KELT-1b only the second sub-stellar object to have a measured offset in its TESS phase curve, after WASP-100b \citep{jansen2020,wong2020}, though the relatively large uncertainty on KELT-1b's phase offset makes detailed comparisons difficult.

The measured nightside blackbody brightness temperature in the TESS band is highly uncertain. Nevertheless, it is consistent with the previously published Spitzer \three and \four nightside temperatures \citep{beatty2019}: $1173^{+175}_{-130}$ K and $1053^{+230}_{-161}$~K, respectively. 

One slight difference between the TESS and Spitzer system parameters is the precise value of $R_{BD}/R_{*}$, which is smaller by 0.0098 ($2.7\,\sigma$) in the TESS data. Since the surface gravity of KELT-1b is approximately 30 times higher than that of a typical hot Jupiter, the atmospheric scale height is very small, and the transit transmission signal is thus well below the measurement uncertainty of both observations. Therefore, in principle, the TESS and Spitzer $R_{BD}/R_{*}$ values should match. Though the difference we measure is only suggestive at $2.7\,\sigma$, it may be indicative of inaccuracies in the modeling of stellar limb darkening either here or in \cite{beatty2019}.

\begin{figure*}[t]
\vskip 0.00in
\includegraphics[width=1.0\linewidth,clip]{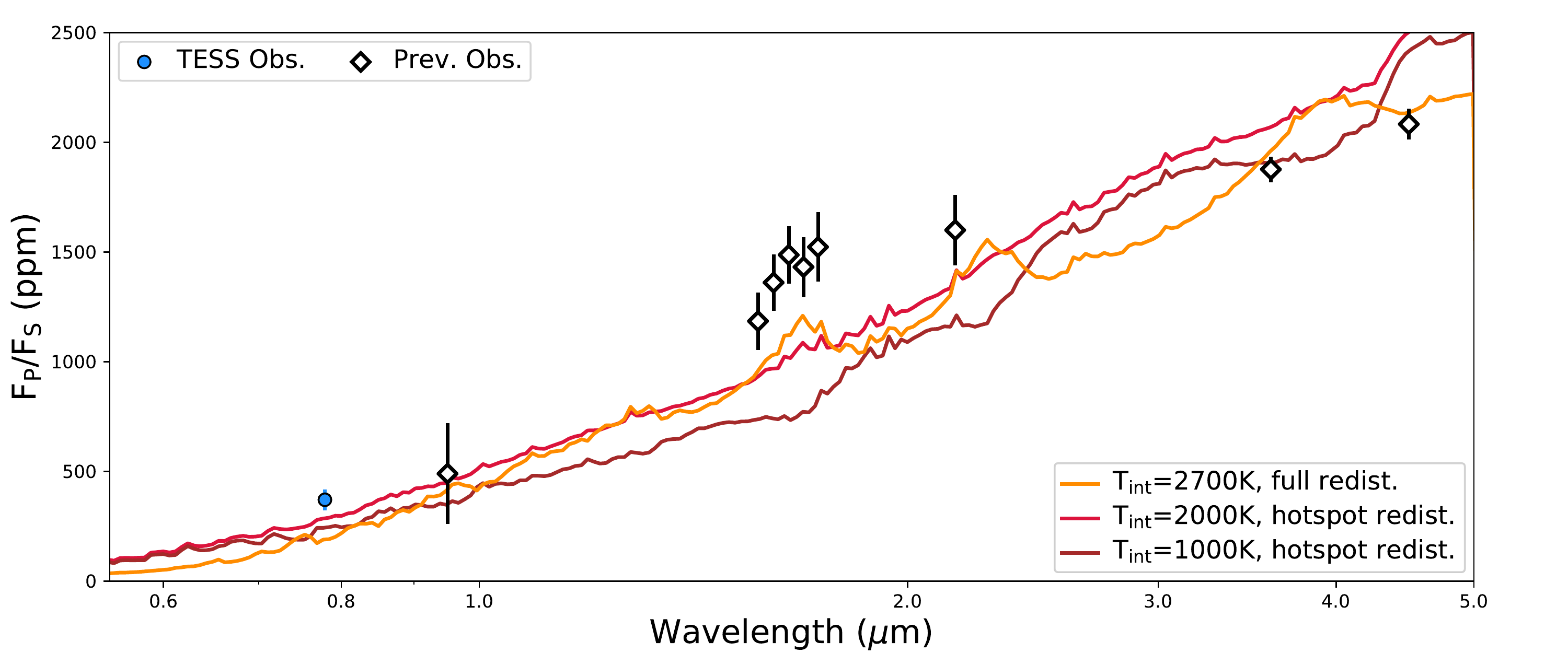}
\vskip -0.0in
\caption{The dayside eclipse spectrum of KELT-1b is poorly fit by 1D equilibrium atmosphere models, likely indicating that there are significant 2D or disequilibrium effects. In particular, all of these cloud-free thermal-only models under-predict the TESS eclipse depth, which suggests that KELT-1b possesses a significant geometric albedo ($\mathrm{A}_\mathrm{g}\sim0.5$) in the TESS bandpass. If this is the case, it is possibly caused by horizontal transport of nightside \citep{beatty2019,keating2019} silicate clouds \citep[$A_s\sim0.6$,][]{sudarsky2000} onto the morning-portion of KELT-1b's dayside -- which would also cause substantial 2D and disequilibrium effects on the eclipse spectrum.}
\label{dayside}
\end{figure*}

The other difference between the TESS and Spitzer phase curves is the blackbody brightness temperature of KELT-1b's dayside in the three bandpasses. \cite{beatty2019} measured brightness temperatures of $3013\pm72$\,K at \three and $2941\pm84$\,K at \fouralt. While the two Spitzer dayside temperatures are mutually consistent, they are both approximately $2.7\,\sigma$ cooler than the $3340\pm110$\,K dayside brightness temperature we measured in the TESS bandpass. Previous ground-based observations of KELT-1b's eclipse spectrum in the $H$-band \citep{beatty2017} have indicated that KELT-1b's dayside spectrum is not a featureless blackbody --- as is common for ultra-hot Jupiters --- so the larger than expected eclipse depth in the TESS bandpass prompted us to consider KELT-1b's eclipse spectrum in more detail.

\subsection{Modeling KELT-1b's Eclipse Spectrum}

In order to interpret the dayside spectrum of KELT-1b, we utilized ScCHIMERA \citep{piskorz2018,arcangeli2018,mansfield2018,wasp103wfc3,gharib-nezhad2019,zalesky2019} to produce a small grid of self-consistent 1D radiative-convective-thermochemical equilibrium models. Since previous work, the model has been upgraded to account for the latest high-temperature ExoMol \citep{tennyson2016} opacities (Gharib-Nezhand et al. in prep) and atomic opacities (Fe I, Fe II, Ca I, Mg I ) relevant to ultra-hot Jupiters \citep[e.g.][]{lothringer2019}. The model machinery has been previously benchmarked and validated against analytic solutions and earlier brown dwarf grid models \citep{saumon2008} in \cite{piskorz2018} and was recently used to interpret the combined WFC3 and Spitzer spectra of the ultra-hot Jupiters WASP-18b \citep{arcangeli2018}, HAT-P-7b \citep{mansfield2018}, and WASP-103b \citep{wasp103wfc3}. 

KELT-1b is a unique object in that while it has a dayside temperature of a typical ultra-hot Jupiter, it has the mass of a brown dwarf, and thus likely has retained a large internal heat flux, or ``internal temperature''. Brown dwarf evolutionary models predict that an isolated field object with KELT-1b's mass and age would have an internal temperature of approximately 900\,K \citep{saumon2008}, but the strong irradiation KELT-1b has received from its primary star has likely severely retarded KELT-1b's cooling \citep{burrows2011}. With this in mind, we explored a range of possible internal temperatures from 100K (similar to a cold-start like planet) to 3500K (similar to an M-dwarf). We also considered several different possible values for heat redistribution: full redistribution from the day- to nightside, dayside-only redistribution, and a minimum ``hotspot'' redistribution \citep[e.g.,][]{wasp18wfc3}. For simplicity we assumed solar composition, though we did change the metallicity slightly with little consequence. We utilized a 6500\,K PHOENIX stellar model for the star KELT-1 \citep[][with 2016 upgrades]{husser2013}.

\begin{figure*}[t]
\centering
\includegraphics[width=\linewidth]{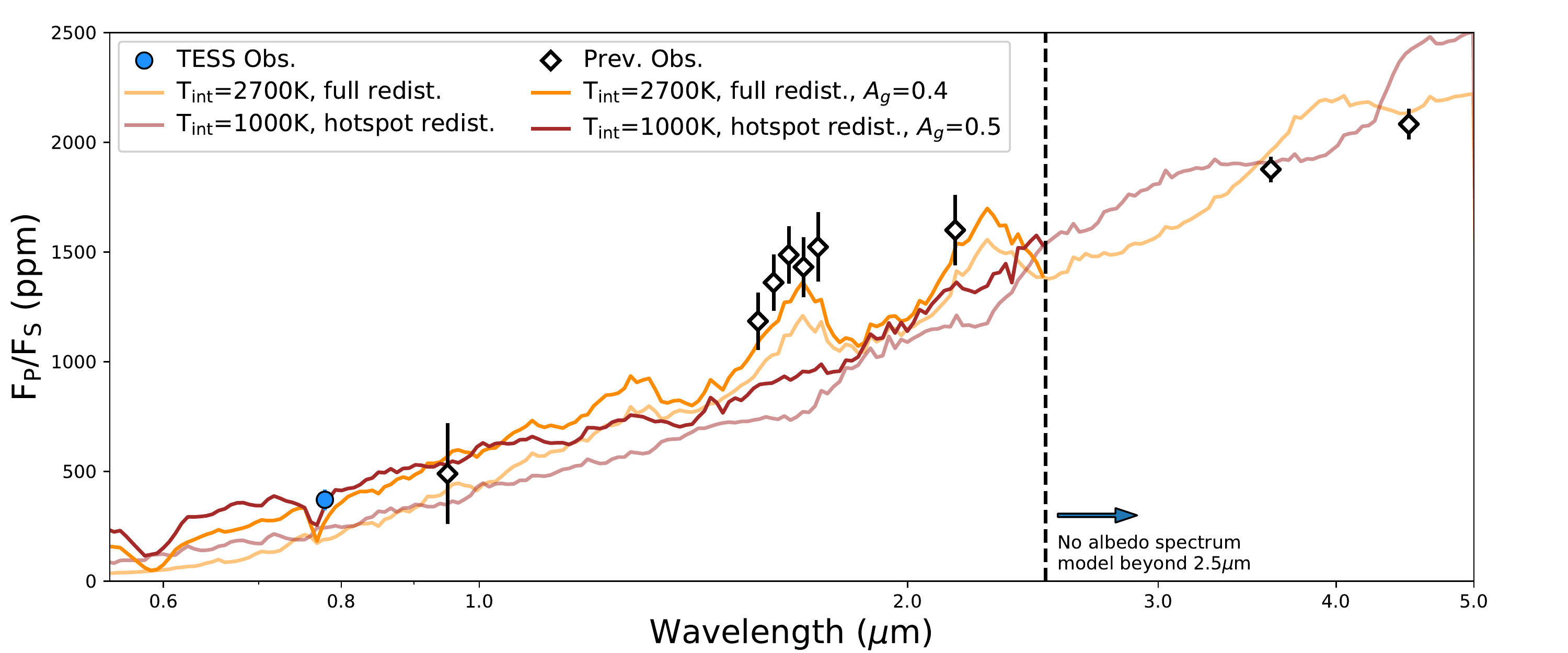}
\caption{Assuming that KELT-1b has a dayside albedo of $\mathrm{A}_\mathrm{g}=0.5$ from silicate clouds \citep{powell2018,gao2020}, and that the $H$- and $Ks$-band data are corrupted due to un-modeled ellipsoidal deformation signals, then the $\mathrm{T}_\mathrm{int}=1000$\,K atmosphere model becomes consistent with the observations at $2.7\,\sigma$. The assumed silicate cloud reflection spectrum is from \cite{sudarsky2000}; note that this model does not extend beyond 2.5\,$\mu$m.}
\label{spec_w_reflect}
\end{figure*}

In addition to our secondary eclipse measurement in the TESS bandpass, KELT-1b's eclipse has been also measured in $z^\prime$ \citep{beatty2014}, $H$-band \citep{beatty2017}, $K_s$-band \citep{croll2015}, and from space by Spitzer at \three and \four \citep{beatty2019}. None of the atmosphere models that we generated provided a good fit to this combined dataset. \emph{A priori} one might have expected results close to the $\mathrm{T}_\mathrm{int}=1000$\,K, ``hotspot'' redistribution, model (Figure \ref{dayside}), since this matches the approximate interior luminosity KELT-1b would have if it were a field object \citep{saumon2008}. However, while this model does fit the Spitzer eclipse depths, it significantly undershoots the ground-based data in the $H$-band and $Ks$-band, as well as the TESS eclipse depth.

The two best fitting models were the $\mathrm{T}_\mathrm{int}=2700$\,K with full heat redistribution model, and the $\mathrm{T}_\mathrm{int}=2000$\,K model with minimum hotspot redistribution (Figure \ref{dayside}) -- but both were still rejected $>6\,\sigma$. The poor fit to these models is primarily caused by the TESS and the ground-based $H$-band observations, which lie significantly above the model predictions. Note that both of these models would be consistent with the observed phase variation even at $\mathrm{T}_\mathrm{int}=2700$\,K, since the observed nightside brightness temperature is $1880^{+680}_{-1210}$\,K.

One possible contributor to the discrepancy between the $H$- and $Ks$-band eclipse depths and our models is the presence of unmodeled ellipsoidal deformation in those measurements. Both the characteristic ellipsoidal deformation and the secondary eclipse reach their respective minima at superior conjunction, so there is a fundamental trade-off between the ellipsoidal distortion amplitude and the secondary eclipse depth. In our fits, this is manifested in a positive correlation between baseline flux of the brown dwarf $F_{0}$ and the amplitude $A_{ellip}$. By not accounting for the presence of ellipsoidal deformation, the eclipse depths reported in \citet{beatty2017} and \cite{croll2015} may be somewhat overestimated, thereby causing the incompatibility between those $H$- and $Ks$-band measurements with the atmospheric models in Figure~\ref{dayside}.

More extensive atmospheric modeling of KELT-1b's dayside spectrum may provide an improved fit to the data. The ScCHIMERA models shown in Figure \ref{dayside} are cloud-free 1D models that assume equilibrium chemistry, but it is likely that even for a cloud-free atmosphere the dayside of KELT-1b shows significant disequilibrium effects caused by the horizontal transport of the cooler nightside atmosphere onto the dayside. 

The high surface gravity of KELT-1b -- as compared to a typical hot Jupiter -- may make this horizontal chemical disequilibrium more noticeable, since KELT-1b's correspondingly high photospheric pressure inhibits most of the molecular dissociation expected on hot Jupiters' daysides \citep{arcangeli2018}. As an example, only 20\% of the water on KELT-1b's dayside should be dissociated, much less than the 99\% water dissociation rate of a hot Jupiter like WASP-103b \citep{parmentier2018}. There should therefore be significant amounts of water present in KELT-1b's atmosphere, and the evolution of water absorption features (and other molecular absorption) across KELT-1b's dayside should change significantly as the atmosphere heats up. The effect of changing spectral emission with longitude is not included in the 1D ScCHIMERA models. 

\subsection{A Possibly High Albedo and Morning Clouds}

While the presence of unmodeled ellipsoidal deformation may resolve the discrepancy with the $H$-band points, the higher than expected TESS band secondary eclipse requires a separate explanation. One possibility is that KELT-1b has a significant optical (and perhaps NIR) albedo. All of the models in Figure \ref{dayside} consider only the thermal emission from KELT-1b's dayside, with no accounting for a possible reflection signal. If the apparently large eclipse depth in the TESS data is due to unmodeled reflection, it would imply that KELT-1b's dayside has a geometric albedo in the TESS bandpass of $\mathrm{A}_\mathrm{g}\sim0.4$ for the $\mathrm{T}_\mathrm{int}=2700$\,K model, $\mathrm{A}_\mathrm{g}\sim0.3$ for the $\mathrm{T}_\mathrm{int}=2000$\,K model, or $\mathrm{A}_\mathrm{g}\sim0.5$ for the $\mathrm{T}_\mathrm{int}=1000$\,K model.

Though HST/STIS observations of ultra-hot Jupiters have failed to detect any dayside geometric albedo in the blue-optical \citep{bell2017}, recent analyses of TESS phase curves have begun to detect evidence for dayside significant reflection for some hot and ultra-hot Jupiters \citep[e.g.][]{daylan2019,wong2020wasp19,wong2020}. The results from \cite{wong2020} are particularly interesting for the case of KELT-1b, since the marginal correlation between increasing dayside temperature and increasing geometric albedo suggested by that work would predict that KELT-1b's albedo is $\mathrm{A}_\mathrm{g}\sim0.2$ to $\mathrm{A}_\mathrm{g}\sim0.4$-- similar to what we find under this interpretation. On the other hand, the possibly high albedo of KELT-1b disagrees with the other tentative correlation found by \cite{wong2020} --- that the dayside geometric albedo appears to decrease with increasing planetary surface gravity. However, the interplay between surface gravity and albedo in \cite{wong2020} was anchored by what was then the only available high-gravity measurement: WASP-18b \citep{shporer2019}. It is therefore possible that WASP-18b is a low-albedo outlier for some as-yet unknown reason -- though this is contingent on KELT-1b having a significant dayside albedo.

One way for there to be a measurable albedo on KELT-1b's dayside would be if there were reflective clouds present. Though the equilibrium temperature of KELT-1b's dayside is too hot for clouds to exist in a steady-state, it is possible that clouds that form on the nightside may be blown over onto the dayside via horizontal transport in the atmosphere. For wind velocities of a few km s$^{-1}$, the rough cloud lifetimes estimated by \cite{helling2019} at these temperatures indicate that clouds could exist for a significant portion of the local morning, and potentially out to the local noon, before breaking-up on the hot dayside.

Recent analyses of Spitzer phase curves results (including that of KELT-1b) have shown that nightside clouds are likely present on all hot Jupiters \citep{beatty2019,keating2019}. Based on the nearly constant 1100\,K nightside temperature of these planets at both \three and \fouralt, \cite{beatty2019} hypothesised that these were primarily silicate clouds, which agrees with simulations \citep[e.g.][]{powell2018,gao2020}. Notably, high temperature silicate clouds have a spherical albedo of roughly $\mathrm{A}_\mathrm{s}\sim0.6$ across the TESS, $H$-, and $Ks$-bandpasses \citep{sudarsky2000} -- similar to the geometric albedos implied above.

If we make the strong assumption that nightside silicate clouds are transported over to the morning of KELT-1b, such that $\mathrm{A}_\mathrm{g}=0.5$, and we discount the $H$- and $Ks-$band data due to un-modeled ellipsoidal deformation, then the $\mathrm{T}_\mathrm{int}=1000$\,K model becomes consistent with the observations at $2.7\,\sigma$ (Figure \ref{spec_w_reflect}).

In principle we should be able to see an effect from any clouds present during KELT-1b's morning in the offset of the TESS phase curve, with the significant clouds driving the phase offset to the west (i.e., negative values). Such a scenario has been detected on Kepler-7b, Kepler-12b, and Kepler-41b from the analysis of the Kepler phase curves \citep{shporer2015}. We measured an eastward offset of $18.3^\circ\pm7.4^\circ$ for KELT-1b, though the uncertainty on that measurement is such that it is also consistent with an offset of zero, or even a few degrees to the west. More generally, given the opposing effects of superrotating winds on thermal emission and the westward offset from morning terminator clouds, our measured offset is likely some combination of a larger eastward thermal offset being balanced by some amount of reflective clouds in the western hemisphere.

\begin{figure}[t]
\centering
\includegraphics[width=\linewidth]{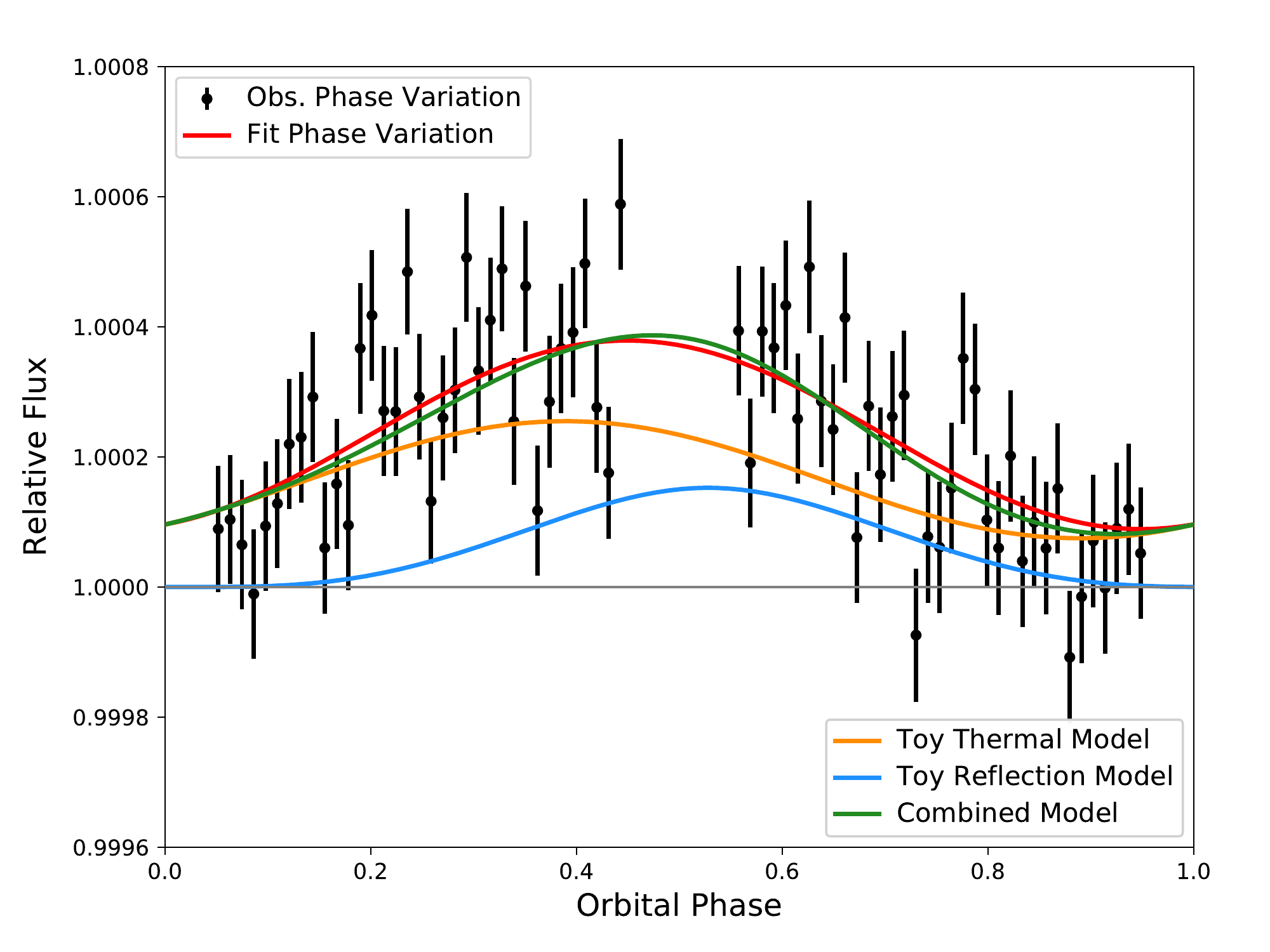}
\caption{If KELT-1b does possess a high dayside albedo, then the observed phase variation is likely the combination of thermal emission and a reflected light. In Section 2.1 we fit the observed phase variation (black points, with the in-transit and in-eclipse data removed) using a single sinusoid that is more representative of a purely thermal emission model (red line). Nevertheless, a combined toy model (green) of a two-component thermal emission (orange) and Lambertian reflected light \citep[blue,][]{esteves2015} variation is also a good fit to the observations.}
\label{toyreflection}
\end{figure}

One aspect regarding a possible high albedo for KELT-1b is that in our fitting process we assumed that the phase variation from KELT-1b is well-modeled by a single sinusoid (Section 2.1), and not by a two-component thermal emission and reflected light model \citep[e.g.,][]{esteves2015}. However, given the size of the data uncertainties relative to the measured phase amplitude (Figures \ref{fig:fit} and \ref{toyreflection}), we considered a more detailed fit to the data to be unwarranted. That being said, as shown in Figure \ref{toyreflection} it is possible to replicate the observed phase variation using a toy model of combined thermal emission and Lambertian reflected light. In this particular case, we assumed that the albedo of the reflected light component is $\mathrm{A}_\mathrm{g}=0.46$, which is consistent with our suggested albedo of $\mathrm{A}_\mathrm{g}\sim0.5$. Note, though, that the combined thermal plus reflected light model shown in Figure \ref{toyreflection} is primarily for illustrative purposes --- to demonstrate that the observed sinusoidal phase variation is also consistent with KELT-1b having a high dayside albedo. A full detailed two-component model fit would need to account for the non-sinusoidal thermal emission seen in the Spitzer phase curve of hot Jupiters \citep{beatty2019} as well as the complex non-Lambertian reflection signal caused by non-uniform dayside clouds. 

\section{Summary and Conclusions}

We have presented our analysis of the TESS full-orbit phase curve of the transiting brown dwarf KELT-1b. We fit the data using a combined model for the secondary eclipse, atmospheric brightness modulation, and the ellipsoidal deformation and Doppler boosting of the primary star. We detrended the TESS data using three different approaches: polynomial detrending, a Gaussian-Process regression, and a Fourier decomposition of the out-of-transit light curve. The results from all three analyses are consistent with each other, and we adopted the values from the polynomial detrending fit as our primary set of results.

We found that KELT-1b shows significant phase variation in the TESS bandpass, with a relatively large phase amplitude of $234^{+43}_{-44}$\,ppm and a secondary eclipse depth of $371^{+47}_{-49}$\,ppm. We also measured a marginal phase offset of $18.3^\circ\pm7.4^\circ$, though given the relatively large uncertainty on the offset measurement this result is also consistent with zero. 

Our measurement of the TESS eclipse and the other measurements of KELT-1b's secondary eclipse at other wavelengths give us a interesting view of the brown dwarf's dayside emission (Figure \ref{dayside}). We modeled this emission using a grid of results from the ScCHIMERA atmosphere models, with KELT-1b's interior temperature and the amount of heat redistribution as the two primary variables affecting the model results. We found three possible fits: one at $\mathrm{T}_\mathrm{int}=2700$\,K with full heat redistribution, one at $\mathrm{T}_\mathrm{int}=2000$\,K with minimum hotspot redistribution, and a model at $\mathrm{T}_\mathrm{int}=1000$\,K also with hotspot redistribution. However, all these models are inconsistent with the combined dayside spectrum at $>6\,\sigma$.

Since the ScCHIMERA models assume 1D, cloud-free, equilibrium chemistry in their calculations, it is possible that the poor fit to the data is caused by these assumptions. Meanwhile, for the higher than expected $H$-band eclipse measurements from \citet{beatty2017}, unmodeled ellipsoidal distortion may explain the discrepant depth measurements relative to our models. In general, KELT-1b may show stronger 2D disequilibrium effects than a typical hot Jupiter. Horizontal transport of cooler nightside atmosphere onto the dayside may introduce a significant source of molecular absorption -- particularly since the high surface gravity of KELT-1b prevents almost all of the dayside molecular dissociation expected on a typical hot Jupiter. In particular, there should be significant water present on KELT-1b's dayside \citep{parmentier2018}.

Given the relatively large eclipse depth we measure in the TESS bandpass, the dayside of KELT-1b may have a significant geometric albedo of $\mathrm{A}_\mathrm{g}\sim0.5$. We hypothesize that this could be caused by the presence of high temperature silicate clouds \citep{gao2020} that form on KELT-1b's nightside and extend significantly onto the morning half of KELT-1b's dayside before breaking up \citep[e.g.][]{helling2019}. If we grant that the $H$- and $Ks$-band eclipse depths are systematically higher due to un-modeled effects from ellipsoidal deformation, then under this hypothesis the $\mathrm{T}_\mathrm{int}=1000$\,K atmosphere model, with a reflection component, is likely the best fit to the data (Figure \ref{spec_w_reflect}). This interior temperature agrees with KELT-1b's expected interior temperature given its mass and age \citep{saumon2008}, and is consistent with the nightside temperatures measured here and in the Spitzer data \citep{beatty2019}.

A high geometric albedo for KELT-1b's dayside would be consistent with the tentative trend found by \cite{wong2020} that hot Jupiters with higher dayside temperatures show higher geometric albedos. Though the steady-state temperatures of these planets' (and KELT-1b's) daysides are too hot for clouds \citep[e.g.][]{parmentier2016, helling2019}, recent ensemble analyses of Spitzer phase curves of hot Jupiters \citep{beatty2019,keating2019} concluded that these planets universally possess nightside clouds. It is possible that clouds forming on the planetary nightsides may be blown over onto the daysides via horizontal transport \citep[e.g.][]{helling2019}. As these clouds break-up on the hot dayside they could provide a significant reflection signal \citep{marley1999,sudarsky2000}. 

If KELT-1b does indeed have a significant dayside albedo caused by partial silicate cloud cover, this would be a strong demonstration of the importance of including horizontal transport and its effect on planetary cloud coverage in 3D atmospheric simulations, as well as the inclusion of cloud effects in 1D modeling of hot Jupiter eclipse spectra. 

\acknowledgements

This paper includes data collected by the TESS mission, which are publicly available from the Mikulski Archive for Space Telescopes. Funding for the TESS mission is provided by the NASA Science Mission Directorate. Resources supporting this work were provided by the NASA High-End Computing (HEC) Program through the NASA Advanced Supercomputing (NAS) Division at Ames Research Center for the production of the SPOC data products.

This work has made use of NASA's Astrophysics Data System, the Exoplanet Orbit Database and the Exoplanet Data Explorer at exoplanets.org \citep{exoplanetsorg}, the Extrasolar Planet Encyclopedia at exoplanet.eu \citep{exoplanetseu}, the SIMBAD database operated at CDS, Strasbourg, France \citep{simbad}, and the VizieR catalog access tool, CDS, Strasbourg, France \citep{vizier}. I.W. is supported by a Heising-Simons \textit{51 Pegasi b} postdoctoral fellowship.

\software{\textsc{batman} \citep{batman}, celerite \citep{celerite}, emcee \citep{emcee}, Astropy \citep{astropy1,astropy2}}

\bibliography{references}

\begin{rotatetable*}
\begin{deluxetable*}{lllll}
\label{tab:results}
\tablecaption{Fit Median Values and 68\% Confidence Intervals}
\tablehead{\colhead{~~~Parameter} & \colhead{Description and Units} & \colhead{\bm{$\mathrm{Polynomial}$}} & \colhead{Gaussian Process} & \colhead{Out-of-Transit}}
\startdata               
\sidehead{System Parameters\tablenotemark{a}:}
~~~$T_C$\dotfill &Transit time (\bjdtdb)\dotfill        & $\bm{2457306.97624^{+0.00028}_{-0.00027}}$ & $2457306.97620\pm0.00027$ & ---\\
~~~$P$\dotfill &Orbital period (days)\dotfill & $\bm{1.21749394\pm2.5\times10^{-7}}$ & $1.2174937\pm2.5\times10^{-7}$ & ---\\
~~~$\cos{i}$\dotfill & Cosine of inclination\dotfill    & $\bm{0.053^{+0.011}_{-0.010}}$ & $0.056\pm0.011$ & ---\\
~~~$R_{BD}/R_{*}$\dotfill &Radius ratio\dotfill         & $\bm{0.07612\pm0.00021}$ & $0.07645\pm0.00025$ & ---\\
~~~$a/R_{*}$\dotfill &Semi-major axis in stellar radii\dotfill & $\bm{3.639_{-0.028}^{+0.025}}$ & $3.65\pm0.03$ & ---\\
~~~$M_{BD}/M_{*}$\tablenotemark{b}\dotfill &Mass ratio\dotfill          & \textbf{---} & $0.0175\pm0.001$ & ---\\[1ex]
\hline \\ [-5ex]
\sidehead{Phase Curve Parameters:}
~~~$F_{0}$\dotfill & Phase baseline (ppm)\dotfill  & $\bm{234^{+43}_{-44}}$  &  $196\pm50$ & --- \\
~~~$C_{1}$\dotfill & Phase amplitude (ppm)\dotfill & $\bm{145^{+20}_{-21}}$   &  $167\pm25$ & 195\\
~~~$C_{2}$\dotfill & Phase offset (deg.)\dotfill   & $\bm{18.3\pm7.4}$ &  $16.9\pm7.1$ & 18\\
~~~$A_{ellip}$\tablenotemark{b}\dotfill & Ellipsoidal def. amplitude (ppm)\dotfill & $\bm{399\pm19}$ & $397\pm20$ & 415\\
~~~$A_{Dopp}$\tablenotemark{a}\dotfill & Dopp. beaming amplitude (ppm)\dotfill &$\bm{41.1\pm1.0}$ & $41\pm0.3$ & 41\\[1ex]
\hline \\ [-5ex]
\sidehead{Derived Parameters:}
~~~$\delta$\dotfill & Secondary eclipse depth (ppm)\dotfill & $\bm{371_{-49}^{+47}}$ &  $355\pm50$ & --- \\
~~~$F_{night}$\dotfill & Nightside flux (ppm)\dotfill       & $\bm{97_{-49}^{+48}}$ &  $37\pm61$ & --- \\
~~~$F_{max}$\dotfill & Phase maximum (ppm)\dotfill          & $\bm{379\pm48}$ & $362\pm50$ & ---\\
~~~$F_{min}$\dotfill & Phase minimum (ppm)\dotfill          & $\bm{89\pm49}$ & $30\pm61$ & ---\\
~~~$T_S$\dotfill &Secondary eclipse time (\bjdtdb)\dotfill & $\bm{2457307.58499^{+0.00028}_{-0.00027}}$ & $2457307.58524\pm 0.00027$ & --- \\
~~~$i$\dotfill & Inclination (deg.)\dotfill & $\bm{87.93_{-0.60}^{+0.64}}$ & $86.7\pm0.6$ & ---\\
~~~$b$\dotfill &Impact parameter\dotfill & $\bm{0.195_{-0.040}^{+0.036}}$ & $0.21\pm0.04$ & ---\\[1ex]
\sidehead{SED-fit and Derived Parameters:}
~~~$R_{*}$\dotfill &Stellar radius (\rsun)\dotfill         & $1.482\pm0.070$ & --- & ---\\
~~~$R_{BD}$\dotfill &Brown dwarf radius (\rj)\dotfill         & $1.105\pm0.051$ & --- & --- \\
\enddata

\tablenotetext{a}{Constrained by Gaussian priors (see Table~\ref{tab:priors} and Section~\ref{section:fitmodel}).}
\tablenotetext{b}{In the fit using polynomial detrending (Section~\ref{section:polyfit}), the ellipsoidal deformation amplitude $A_{ellip}$ is a free parameter, from which the system mass ratio $M_{BD}/M_{*}$ is subsequently derived; vice versa for the case of GP regression (Section~\ref{section:gp}).}
\end{deluxetable*}
\end{rotatetable*}

\end{document}